\documentclass{mem}
\usepackage{natbib}\usepackage{txfonts}\usepackage{balance}
\usepackage{flushend}
\usepackage{graphicx}
\usepackage{xcolor}
\usepackage[a4paper,breaklinks,pdftex]{hyperref}
\idline{0}{0}

\begin{document}

\title{Machine learning approach to the detection of point sources in maps of the CMB temperature anisotropies}

\author{
P. \,Diego-Palazuelos\inst{1,2},
R. B. \,Barreiro\inst{1},
P. \,Vielva\inst{1},
D. \,Balb\'as\inst{3},
M. \,L\'opez-Caniego\inst{4},
D. \,Herranz\inst{1,2},
\and B. \, Casaponsa\inst{1}
}
 
\institute{
Instituto de F\'isica de Cantabria (CSIC-UC), Santander, Spain
\and
Departamento de F\'isica Moderna, Universidad de Cantabria, Santander, Spain
\and
IMDEA Software Institute, Madrid, Spain 
\and
Aurora Technology for the European Space Agency, Madrid, Spain
\\
\email{diegop@ifca.unican.es}
}

\authorrunning{Diego-Palazuelos et al. }
\titlerunning{Detection of point sources in CMB temperature maps}

\date{Received: Day Month Year; Accepted: Day Month Year}

\abstract{
We propose a machine learning approach to the blind detection of extragalactic point sources on maps of the temperature anisotropies of the cosmic microwave background. Using realistic simulations of the microwave sky as seen by \textit{Planck}, we train a convolutional neural network (CNN) that solves source detection as an image segmentation problem. We divide the sky into regions of progressively increasing Galactic foreground intensity and independently train specialized CNNs for each region. This strategy leads to promising levels of completeness and reliability, with our CNN substantially outperforming traditional detection methods like the matched filter in regions close to the Galactic plane.

\keywords{Cosmic Microwave Background -- Point Source Detection -- Convolutional Neural Networks -- Image Segmentation }
}
\maketitle{}

\section{Introduction}

To extract cosmological information from cosmic microwave background (CMB) observations, we must first separate its signal from the rest of Galactic and extragalactic emissions~\citep{{component_separation}}. Therefore, although they are an interesting subject of study on their own~\citep{PS_CORE}, extragalactic sources emitting in the microwave range (e.g., radio-loud active galactic nuclei or dusty galaxies) constitute a contaminant that must be removed to extract information from the small angular scales of CMB data~\citep{{PS_contamination}}.

Traditionally, algorithms for the detection of compact sources have combined wavelet or linear filters defined to maximize the signal-to-noise ratio of sources with respect to the background with a thresholding detection criterion~\citep{PS_detection}. As an alternative, we propose a machine learning approach to the blind detection of extragalactic point sources on maps of the temperature anisotropies of the CMB. We  treat source detection as an image segmentation problem and design a convolutional neural network (CNN) that successfully solves it while maintaining a simple autoencoder architecture. Note that with this approach we are decoupling the process of source localization from flux estimation. As it does not need to reconstruct the flux of sources, our CNN can detect sources down to lower fluxes and inmersed in more complex astrophysical backgrounds than algorithms based on unbiased flux estimators like the matched filter (MF)~\citep{MF}.
 
\section{Methodology}\label{sec:methodology}

We train and test our CNN using realistic simulations of the microwave sky as seen by the 143~GHz frequency band of the \textit{Planck} mission~\citep{Planck_overview}. These mock observations contain CMB realizations according to \textit{Planck}’s best-fit cosmological model~\citep{Planck_cosmology}, realistic simulations of instrumental noise~\citep{Planck_overview}, and Galactic and extragalactic foregrounds. We use the \textit{Planck Sky Model}\footnote{\url{https://pla.esac.esa.int}}~\citep{PSM} to simulate the synchrotron, thermal dust, spinning dust, and free-free diffuse Galactic emissions, and the extragalactic thermal and kinetic Sunyaev-Zeldovich effects. We generate extragalactic point sources following the number counts per flux density models of \cite{Tucci2011}. Taking the \cite{PCCS2} catalog as reference, we distinguish between detectable (labeled) sources and a background of faint undetectable (unlabeled) sources with fluxes below 177~mJy.

We project the sky into two-dimensional $3.63^\circ\times3.63^\circ$ patches and, based on their mean Galactic foreground intensity, divide them into partially overlapping regions of faint, medium, and bright foreground emission. These regions approximately cover a 65\%, 40\%, and 25\% fraction of the sky, respectively. Inside each region, patches are separated into alternating rings of equal longitude to evenly divided them into training and validation sets that do not overlap between them.

\begin{figure}[]
\resizebox{\hsize}{!}{\includegraphics[clip=true]{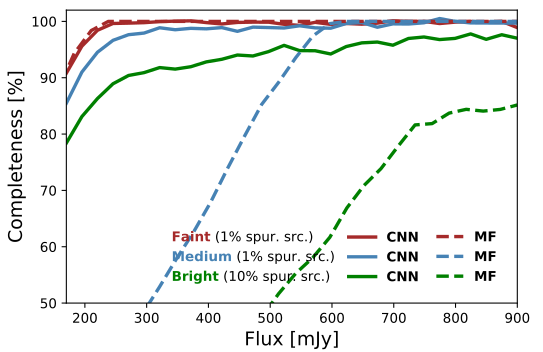}}
\caption{\footnotesize Percentage of detected sources above a given flux using the MF (dashed) and our CNN (solid). Results are shown for the regions of faint (red), medium (blue), and bright (green) Galactic foreground emission, indicating the percentage of spurious detections allowed in each case.}
\label{results}
\end{figure}

We use supervised learning with binary labels to train independent CNNs that specialize in each region. However, the higher complexity of foregrounds and the reduced size of the training sample lead to over-fitting on the region of brightest foreground emission. Preliminary results show that the generalization ability of the CNN trained for bright foreground emission improves when increasing the size of the training set with rotated and re-scaled patches from other regions of the sky and applying regularization techniques such as dropout~\citep{dropout}. After implementing these procedures, the results presented in Section~\ref{sec:results} do not show any significant signs of over-fitting.

\section{Results}\label{sec:results}

In Figure~\ref{results}, we show the preliminary results that we obtain when applying the CNN to the validation set defined in Section~\ref{sec:methodology} and compare them with those obtained using traditional methods like the MF. Both methods perform similarly well in regions far from the Galactic plane where the CMB dominates over the foreground emission (red lines).

Once we move to regions of the sky dominated by Galactic emission, the MF is outperformed by the CNN as it is no longer able to characterize the statistics of the background. Compared to the MF, our CNN reduces the flux at which a 90\% completeness is reached from 514~mJy to 195~mJy  (1\% of spurious detections allowed) in the region of intermediate foreground intensity (blue lines). It is inside the Galactic plane (green lines) where our CNN would provide a substantial improvement on the completeness of current catalogs, achieving a 90\% completeness at 270~mJy while yielding only a 10\% of spurious detections. Under the same conditions, the flux threshold of the MF is at 1013~mJy.

\section{Conclusions and future work}

Although dropout and data augmentation can satisfactory solve our over-fitting problem, we would like to further improve the robustness of our CNN by exploring different options. E.g., increasing the number of training samples for all regions by reducing the size of patches, or applying transfer learning~\citep{transfer_learning} to progressively specialize the CNN towards regions of brighter foreground emission instead of training independent CNNs for each region.

Once over-fitting is fully under control, we will be ready to extend the training to other frequencies. Our preliminary results lead us to believe that, once applied to data, the CNN will produce deeper and more complete catalogs of extragalactic sources.

\begin{acknowledgements}
PDP, RBB, PV, and DH acknowledge funding from the Unidad de Excelencia Mar\'ia de Maeztu (MDM-2017-0765). PDP, RBB, and PV thank the Spanish Agencia Estatal de Investigación (AEI, MICIU) for the financial support provided under the project PID2019-110610RB-C21, and DH for that under the project PGC2018-101814-B-I00. PDP also acknowledges funding from the \textit{Formaci\'on del Profesorado Universitario} program of the Spanish Ministerio de Ciencia, Innovaci\'on y Universidades. We acknowledge the use of the \textit{Planck Sky Model}, \texttt{Tensorflow}, \texttt{Keras}, \texttt{HEALPix}, \texttt{Scipy}, \texttt{skimage}, \texttt{Matplotlib}, and \texttt{NumPy}.
\end{acknowledgements}

\end{document}